\documentclass[12pt]{article}

\usepackage[utf8]{inputenc}
\usepackage{tabularx,booktabs,colortbl}
\usepackage{amsmath,array,graphicx}
\usepackage{textcomp}
\usepackage{gensymb}
\usepackage{float}
\usepackage{graphicx}
\usepackage{textgreek}
\usepackage{multicol}
\usepackage{makecell}
\usepackage[labelfont=bf]{caption}
\usepackage[textfont=it]{caption}
\usepackage[margin=0.75in]{geometry}
\usepackage[superscript]{cite}
\usepackage{tabularx}
\title{Active learning and molecular dynamics simulations to find high melting temperature alloys}

\author{David E. Farache$^{a,*}$, Juan C. Verduzco$^{a,*}$, Zachary D. McClure$^a$,\\ Saaketh Desai$^b$, and Alejandro Strachan$^a$}

\date{
$^a$ School of Materials Engineering and Birck Nanotechnology Center \\
Purdue University, West Lafayette, Indiana, 47907 USA \\

$^b$ Center for Integrated Nanotechnologies \\ 
Sandia National Laboratories, Albuquerque, New Mexico, 87123 USA \\

* Authors contributed equally to this work. 
}

\begin{document}

\maketitle

\begin{abstract}

Active learning (AL) can drastically accelerate materials discovery; its power has been shown in various classes of 
materials and target properties. Prior efforts have used machine learning models for the optimal selection of 
physical experiments or physics-based simulations. However, the latter efforts have been mostly limited to the 
use of electronic structure calculations and properties that can be obtained at the unit cell level and with 
negligible noise. We couple AL with molecular dynamics simulations to identify multiple principal component alloys 
(MPCAs) with high melting temperatures. Building on cloud computing services through nanoHUB, we present a fully 
autonomous workflow for the efficient exploration of the high dimensional compositional space of MPCAs. We characterize 
how uncertainties arising from the stochastic nature of the simulations and the acquisition functions used to select 
simulations affect the convergence of the approach. Interestingly, we find that relatively short simulations with 
significant uncertainties can be used to efficiently find the desired alloys as the random forest models used for AL 
average out fluctuations.

\end{abstract}


\section{Introduction}
\label{introduction_section}

\begin{figure}[H]
     \centering
     \includegraphics[width=0.8\textwidth]{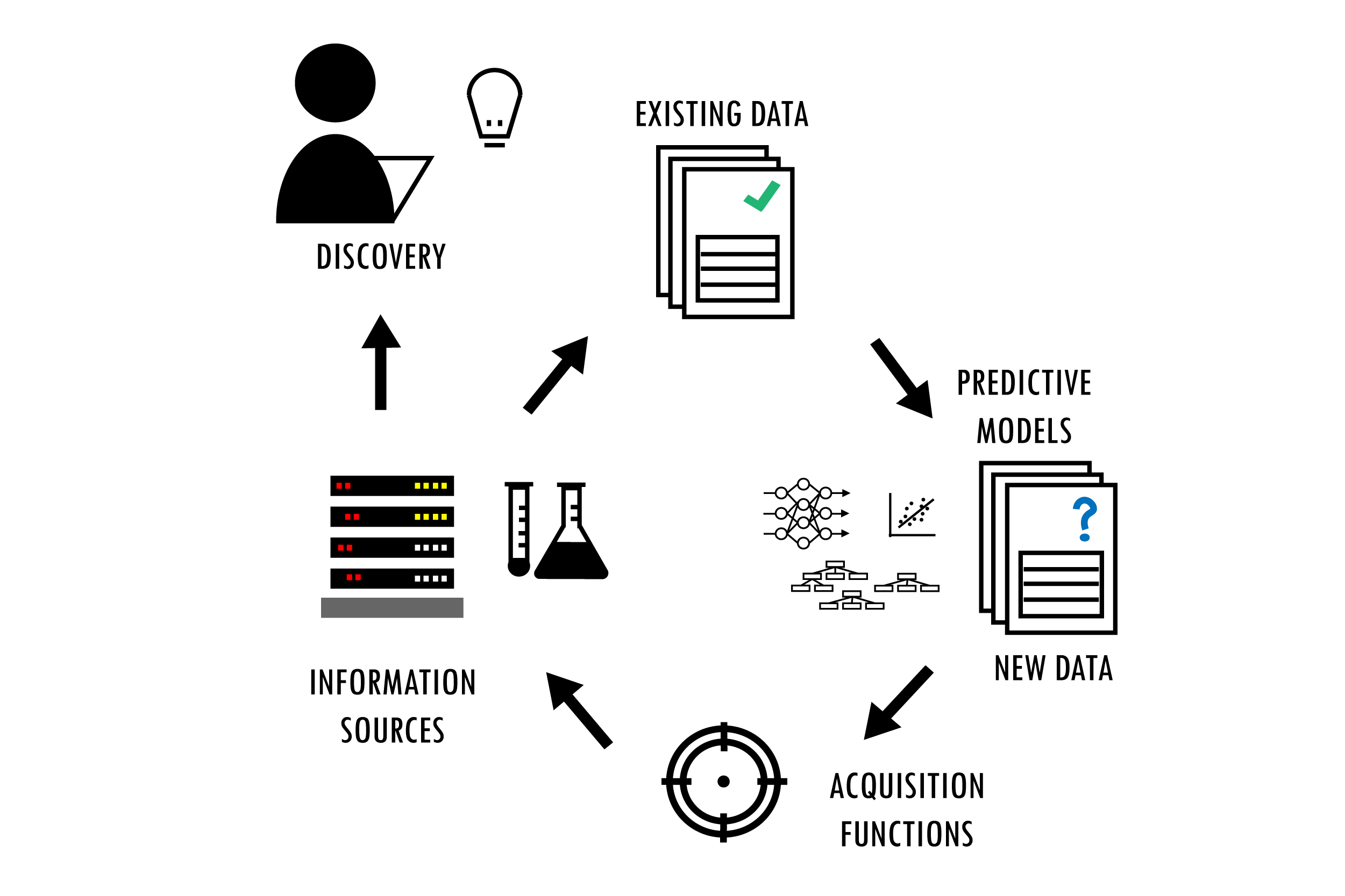}
     \caption{Schematic representation of the AL iterative optimization. Starting from existing data, a predictive model is trained with machine learning. The model is then applied to all candidate materials in the design space to assess their performance, including uncertainties. An acquisition function is used to select the next material(s) to be tested (either via experiments or simulations). 
     If the new material does not satisfy the design conditions, it is added back to the existing data set and the cycle re-initiated.}
     \label{activelearning}
\end{figure}

The combination of experiments, physics-based simulations, and data science tools has been shown to have the potential 
to accelerate discovery of novel materials with optimized properties \cite{eyke2020iterative, zhao2018active, kusne2020fly, xue2016accelerated, tran2018active, jain2013commentary, seko2015prediction, desai2020introduction, balachandran2016adaptive, kim2019active, ling2017high, verduzco2021active}.
Active learning (AL), a subset of machine learning in which models learn dynamically, has gathered significant interest, both from the point 
of basic science \cite{national2011materials} and for commercial applications \cite{wise2019implementation}.
AL models analyze existing data and formulate queries to acquire additional information towards a design goal. AL workflows start with a 
model trained with an initial set of data and evaluate the expected gain towards an objective function of all possible new experiments within 
a design space \cite{settles_synthesis_2011}.
Top candidates are characterized (e.g. by performing an experiment) and the outcome is added to the existing dataset. Given this additional 
information, a new cycle is started. With this iterative process, illustrated in Figure \ref{activelearning}, the model becomes more 
accurate in regions of interest within the design space. In order to identify the next best query, AL uses selection strategies known as 
information acquisition functions. These strategies differ from each other in the relative balance between exploitation and exploration 
included in their mathematical formulation. Exploitation favors cases expected to maximize the objective function while exploration focuses 
on areas of high uncertainty. 

AL has been used for a wide range of applications from natural language processing \cite{shen2017deep}, 
reaction screening for pharmaceutical applications \cite{eyke2020iterative}, and multiscale modeling 
\cite{zhao2018active,yoo2021neural}.  
In materials science, AL has been used to accelerate discovery of materials with desired properties by coupling it with
experiments \cite{kusne2014fly, xue2016accelerated, nikolaev2016autonomy, ren2018accelerated, noack2019kriging, kiyohara2016acceleration} and physics-based simulations \cite{tran2018active, jain2013commentary, seko2015prediction}. 
In addition, AL workflows paired with existing closed data sets has shown the ability of these models to reduce the number of queries needed to identify the best candidate 
\cite{balachandran2016adaptive, kim2019active, ling2017high, verduzco2021active}.

Kusne et al. \cite{kusne2020fly} developed a closed-loop autonomous algorithm for materials discovery and used it to explore the 
Ge-Sb-Te ternary system in search for an optimal phase-change material with the highest optical bandgap difference between phases. 
With an iterative approach using experimental ellipsometry data, they identified a new composition, Ge$_4$Sb$_6$Te$_7$, with nearly 
three times the optical bandgap of Ge$_2$Sb$_2$Te$_5$, a material used in random-access memory devices.
Xue et al. \cite{xue2016accelerated} demonstrated an adaptive design strategy for the search of NiTi-based shape memory alloys with 
low thermal hysteresis. Starting with 26 materials synthesized their optimization algorithm was used to screen a space of $\sim$800,000 
candidate materials. This screening resulted in 36 alloys queried and tested using differential scanning calorimetry (DSC) with several 
showing lower thermal hysteresis than the starting training set. For additional examples, see Refs. \cite{nikolaev2016autonomy, ren2018accelerated, noack2019kriging, kiyohara2016acceleration}.

Coupling AL workflows with physics-based simulations can further reduce the costs and time required to discover new materials by 
minimizing the number of experimental trials required. Tran and Ulissi \cite{tran2018active} explored a high dimensional space of 
density functional theory (DFT) results on intermetallics in search for optimized catalyst materials for CO$_2$ reduction. Their 
approach produced a set of 54 promising materials from $\sim$1499 candidates in the Materials Project \cite{jain2013commentary}. Out 
of these materials, some have been since confirmed experimentally. Seko et al. \cite{seko2015prediction} reported a virtual screening 
of 54,779 materials looking for compounds of low lattice thermal conductivity. They used AL to select appropriate descriptors for 
a Gaussian process regression (GPR) model used to get information on the large dataset. Their screening yielded 221 materials with 
very low conductivity. After some constraints were imposed, 19 of them were characterized using DFT calculations for possible 
applications as thermoelectric materials.

Work involving the combination of AL or similar optimization methods and materials simulations has been mostly limited to DFT 
calculations and to properties that can be obtained from relatively small simulation cells without noise. 
While DFT often offers a good balance between accuracy and computational efficiency, it remains limited to a subset of 
materials properties. Fortunately, materials models across scales are available \cite{van2020roadmap} and significant progress in 
their coupling across scales has occurred over the last few decades. Examples range from crystal plasticity models 
\cite{cuitino2001multiscale, kim2019active} to interatomic potentials from ab-initio simulations \cite{strachan1999phase,van2001reaxff, bartok2010gaussian,thompson2015spectral, behler2007generalized,yoo2021neural}. Physics-based materials models across scales open the 
possibility of significantly expanding the reach of AL approaches. 

Molecular dynamics (MD) is an important rung in multiscale modeling that connects {\it ab initio} physics to the meso- and
macro-scales. MD has been paired with optimization methods like reinforcement learning, including Monte Carlo tree search 
(MCTS) \cite{m2017mdts, dieb2019monte}. For example, Patra et al. \cite{patra2020accelerating} paired MCTS with MD to find a 
copolymer compatibilizer, and Loeffler et al. \cite{loeffler2021reinforcement} applied the methods to study defect 
structures in metal dichalcogenides. 

In this work, we couple AL with MD simulations to find multiple principal component alloys (MPCAs) with high predicted melting temperature for a model CrCoCuFeNi system. The main objective of this paper is to demonstrate the potential of coupling of AL 
workflows with MD simulations and characterizing the effect of the stochastic and noisy nature of MD results on AL workflows.
As discussed in Section \ref{MDSims}, given the uncertainties associated with the predictions of melting temperature with the current potential, the composition with the highest melting temperature predicted by the model may not match experiments.
This paper is organized as follows: 
Section \ref{Methods_section} introduces the overall approach, design space, and simulation details.  
Section \ref{activelearning_section} describes the AL runs exploring the effects of (a) different information acquisition functions 
and (b) different MD simulation times. It also discusses how uncertainty of the model evolves as the active learning workflow progresses. 
In Section \ref{conclusion_section} we summarize and draw final conclusions of the paper. Finally, in Section \ref{availability_section} we discuss availability of our models, datasets and workflows through nanoHUB.  


\section{Methods}
\label{Methods_section}



\subsection{Problem statement}

MPCAs are a new class of materials, which includes high entropy and complex concentrated alloys, where four of more elements are combined in nearly equal atomic percentages. Interest in these materials has grown due to several desirable properties. Properties such as high strength at elevated temperatures, radiation resistance \cite{gorsse2018high}, and high melting temperatures make some MPCAs attractive for high temperature and extreme applications \cite{lu2014promising, xu2019design, praveen2018high}. Experimental determination of this melting temperature is challenging because of their complex processing \cite{SENKOV2010refract} protocols, chemical reactions, and phase separation \cite{wang2016liquid, he2017phase}. For these reasons, using computational methods to determine the melting temperature of these alloys is highly desirable. Fortunately, melting temperature can be accurately obtained from first principles, and the first calculations for MPCAs are emerging \cite{hong2021theoretical}.

The high-dimensional space of possible compositions prevents brute-force approaches even for relatively cheap simulations. Therefore, an efficient way to explore this space is essential. In this work, we pair AL with MD simulations to find MPCAs with the 
highest possible temperature in a model system. AL in the context of materials design seeks to reduce the number of experiments 
needed to find an optimal candidate out of a pool of untested contenders in an unexplored space. Within the AL scheme, predictions 
of the surrogate machine learning model, along with sample-wise uncertainties, are used to identify the next candidate to be tested, 
via acquisition functions. In this work, we make use of the FUELS (Forests with Uncertainty Estimates for Learning Sequentially) framework 
proposed by Ling et al. \cite{ling2017high} based on random forests and a supervised machine learning algorithm.

\subsubsection{Initial and candidate space}
\label{SearchSpace}

We explore 5-component FCC alloys incorporating Cr, Co, Cu, Fe, and Ni. Iterative AL models require an initial subset of entries to be 
evaluated to serve as the prior for the ML model. Our initial set of alloys was selected to be around the equiatomic composition 
Cr$_{20}$Co$_{20}$Cu$_{20}$Fe$_{20}$Ni$_{20}$. Alloys included in this initial set deviate by 10 at.\% from this composition, such that each element composes 10 at.\% to 30 at.\% of the alloy. Using steps of 10 at. \%,  we get 39 data points. Melting temperatures were calculated via MD simulations, as described below, using the two-phase coexistence method. This initial set was the starting point for all of our AL runs.

For our exploration space, we relaxed the constraints from the initial set to allow more deviations from the equiatomic such that any of 
the elements is between 0 at.\% and 50 at.\%. This allows for exploration of binaries, ternaries and quaternaries and ultimately, using 
steps of 10 at. \%, results in 554 data points that are not included in our initial subset. We recognize that this grid spacing is coarse, 
but it is appropriate for an initial exploration. Our iterative AL model will start with training on our initial space, and the candidate 
space will serve as the pool from which it will be drawing candidates.

\subsection{MD simulations of melting}
\label{MDSims}

To predict the melting temperature of FCC MPCAs, we made use of MD simulations and the two-phase coexistence method \cite{morris1994melting, morris2002melting}. To describe atomic 
interactions, we used a many-body, embedded-atom-method interatomic potential developed by Farkas and Caro \cite{farkas2018model}. The potential 
was designed for FCC Cr-Co-Cu-Fe-Ni alloys with near equiatomic compositions. Its parameterization focused on reproduction of elastic constants, vacancy formation energy, stacking fault energy, surface energies and relative phase stability to ensure that FCC is the stable structure for all five components. However, this potential has not been trained or tested for melting temperature calculations.

Several techniques exist to compute the melting temperature of materials using MD simulations. The most direct approach involves heating a crystalline sample until it melts and then cooling the melt until it recrystallizes. This results in overheating and 
undercooling, requiring a-posteriori corrections \cite{luo2003maximum}. To avoid these issues, we predict melting temperatures 
using a phase coexistence method \cite{morris1994melting, morris2002melting}. In this method, a temperature at which the crystal 
and liquid phases of the material of interest coexist is established.  

\begin{figure}[H]
     \centering
     \includegraphics[width=0.5\textwidth]{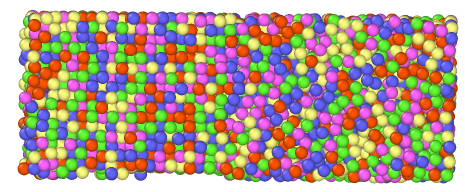}
     \caption{MD snapshot of the simulation cell divided into a liquid and solid region for an equiatomic alloy of Cr,Co,Cu,Fe and Ni. Each color represents a different element.}
     \label{OVITO+SC}
\end{figure}

The first step is to generate an initial sample with a liquid and solid phase in contact, we achieve this with the
use of two thermostats during the sample preparation step. We start from the four-atom FCC unit cell with a lattice parameter 
of 3.56 \AA{} and replicate it 8 x 8 x 18 to create a periodic simulation cell. The third direction (z) is normal to the 
solid-liquid interface. Atom types are assigned based on the desired composition randomly to each lattice site. 
The entire structure is relaxed via energy minimization with respect to cell parameters and atomic positions. 
To create initial liquid/solid structure the cell is then divided in halves along z and two thermostats are 
used to create the two phases. The two regions are equilibrated at their initial temperatures for 10 ps using isothermal, isobaric 
MD simulations (NPT ensemble) with a Berendsen thermostat \cite{berendsen} and Nose-Hoover barostat \cite{n-h}.  
This first step seeks to create an initial cell with both liquid and solid and the two temperatures need to chosen 
appropriately (the temperature of the liquid half should be higher than the melting temperature). 
A snapshot of the system after this initial step is shown in Figure \ref{OVITO+SC}. 
Clearly, this initial sample is not in equilibrium, and we follow the initial step with an isoenthalpic, isobaric (NPH) 
simulation where the liquid and solid regions are allowed to exchange energy freely and come to equilibrium. 
After steady state is achieved and the simulation cell has a uniform temperature, if the sample contains both liquid and solid 
the simulation temperature corresponds to the melting temperature of the system. 
The use of the NPH ensemble allows the temperature of the system to evolve towards the melting temperature. If 
the sample is initially too hot the liquid phase will grow at the expense of the solid, the heat of fusion will automatically cool 
the sample down towards the melting temperature; the reverse happens if the sample is initially too cold.

A successful run requires the coexistence of liquid and solid after steady state has been achieved. These conditions are 
checked by analyzing the atomic structure and the time evolution of the instantaneous temperature of the system. 
Phase fractions of solid and liquid are calculated using the polyhedral template matching (PTM) algorithm from OVITO \cite{larsen2016robust,stukowski2009visualization}. Coexistence is defined as the fraction of liquid being between 35 and 65 at.\% 
and the FCC crystal also between 35 and 65 at.\% in the final snapshot of the simulation. We also analyze thermodynamic data from 
the last 60\% of the simulation to assess steady state, we require the absolute value of the slope of the instantaneous 
temperature vs. time to be less than 1 K/ps. Finally, we also compute the 95\% confidence interval on the temperature calculation. Melting temperatures for our initial set ran with 100ps of simulation time.

\subsection{Uncertainties and noise in the MD predictions}
\label{MD-UQ}

\begin{figure}[H]
     \centering
     \includegraphics[width=0.6\textwidth]{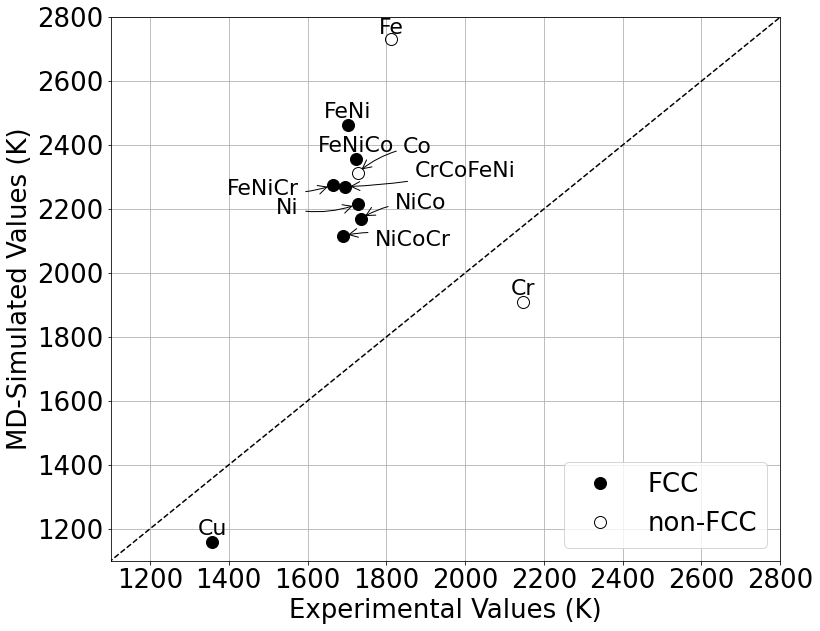}
     \caption{Comparison for MD simulated and experimental melting temperatures for alloys composed of elements included in the potential. Dashed line indicates a match between the values. Experimental values for MPCAs taken from Wu et al \cite{WU2014melt}. Experimental values for single-elements taken from NIST \cite{valencia2013thermophysical}.Filled symbols represent alloys reported as FCC crystal structure, open symbols represent non-FCC crystal structures.}
     \label{single_phase_parity}
\end{figure}

The description of interatomic forces is at the heart of MD simulations and determines the predicted melting temperatures. The 
potential used in this work was not developed or widely tested for melting temperature calculations and we are unaware of any potential that
has. Thus, our focus is on understanding the use of MD in ML workflows and not in accurate predictions of melting temperatures. 
Nevertheless, for completeness, Figure \ref{single_phase_parity} compares the predicted melting temperatures with experimental 
values \cite{WU2014melt} for various alloys and single-elements \cite{valencia2013thermophysical}. Our simulations assume random FCC alloys since that is what the AL workflow uses, this does not match the experiments in all cases. 
We included experimental melting temperatures for single elements for comparison. All MD simulations all arrange structures as FCC. 
For iron, we show the predicted melting temperature of FCC Fe since the value is relevant for the high-melting temperature
alloys found by the AL workflow. Interestingly, for Cobalt (Co), the MD simulation with this potential achieved coexistence by creating a solid BCC structure rather than FCC. We find that there is a consistent overestimation of melting temperatures, except for Cu and Cr. 

It is important to note that the generation of the initial atomic structure is done stochastically, and temperature is obtained as the time-average of a fluctuating quantity. Thus, the resulting melting temperature from independent runs will vary due to sample-to-sample variability (which can be reduced increasing the simulation cell size) and time averaging of instantaneous temperature (which can be reduced by running longer simulations and also by increasing system size). One of the goals of this paper is to assess what level of noise can be tolerated in an AL workflow.

\begin{figure}[H]
     \centering
     \includegraphics[width=0.75\textwidth]{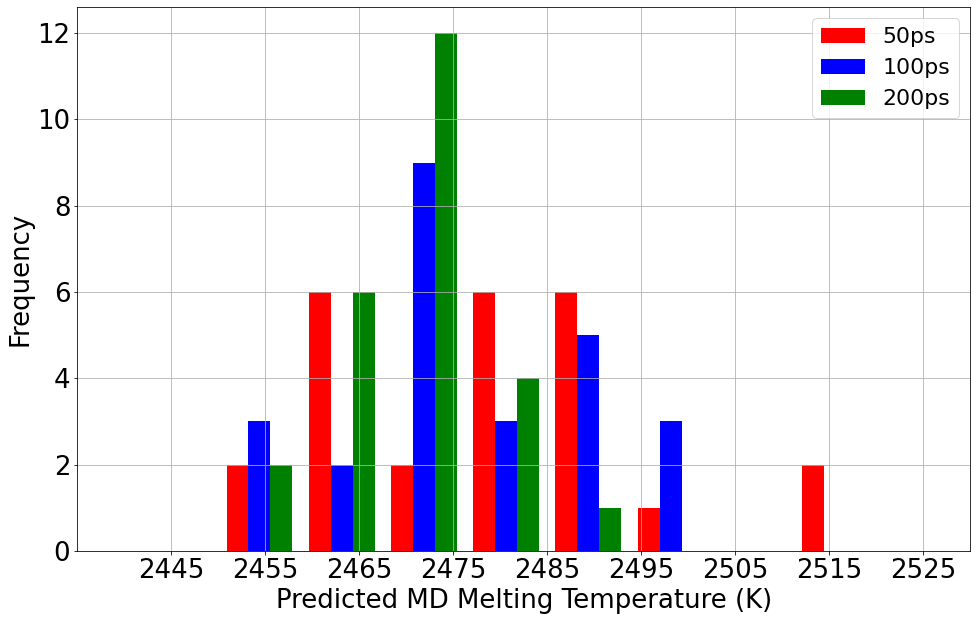}
     \caption{Distributions of MD simulated melting temperatures for a MPCA of 50\% Fe and 50\% Co at different simulation times. Distributions indicate a mean temperature around 2473 K and narrow down as the simulation time increases.}
     \label{MeltingTempDistribution}
\end{figure}

To quantify the variability in predicted melting temperatures, we chose one of our optimal alloy candidates. We performed 36 simulations with the same overall composition but independent atomic configurations and initial velocities. Figure \ref{MeltingTempDistribution} shows the resulting distributions of predicted melting temperatures for the three simulation times used in this study, the inset shows the mean and uncertainty estimates. Figure \ref{MeltingTempDistribution} shows that, as expected, repeated experiments result in mean values relatively independent of the simulation time (2470-2480 K) and reducing the simulation time results in broader distributions (remember that temperature is associated with the time-averaged kinetic energy per degree of freedom).

\begin{table}[H]
\caption{Mean and uncertainty estimates for the distributions in Figure \ref{MeltingTempDistribution} of MD simulated melting temperatures for a MPCA of 50\% Fe and 50\% Co at different simulation times. These distributions were constructed by running 36 simulations with combinations of 6 different atomic arrangements and 6 different atomic velocities initializations.}
\label{histogram_table}
\vskip 0.01 in
\begin{center}
\begin{small}
\begin{sc}
\resizebox{.6\textwidth}{!}{%
\begin{tabular}{lccr}
\toprule
\midrule
\thead{Time Per \\ Simulation (ps)} & \thead{Mean MD Simulated \\ Tm (K)} &  \thead{Standard deviation \\ MD Simulated Tm (K)}\\
\midrule
50 & 2479  & 17 \\
100 & 2476  & 13 \\
200 & 2470 & 8 \\

\bottomrule
\end{tabular}}
\end{sc}
\end{small}
\end{center}
\vskip -0.05in
\end{table}

\subsection{Random forests and uncertainties}
\label{rf_section}

Several regression models, including neural networks, gaussian processes, and random forests have been used in AL workflows. Neural networks have stark advantages in image recognition and other high-dimensionality problems and have seen widespread adoption in a myriad of problems and disciplines. However, uncertainty quantification on their applications for regression and classification is remains an active area of research \cite{tripathy2018deep}. Random forests, on the other hand, are often preferred when dealing with tabulated data with well-defined and limited descriptors \cite{chen2016xgboost}. Random forests also have the ability to yield descriptor importance for interpretability \cite{breiman2001random} and there has been significant work on quantification of uncertainties \cite{ling2017high, efron2012model, wager2014confidence}.

In this work, we selected random forests as uncertainties are critical to make use of information acquisition functions within AL and 
to deal with noisy data present within our simulations. We use the lolopy library \cite{hutchinson2016citrine} to develop RF 
models for melting temperature of MPCAs using composition and derived descriptors as inputs. The models provide an expectation value and 
sample-wise uncertainties that are fed to various acquisition functions to select the next experiment. Uncertainty estimates are obtained 
based on the jackknife after bootstrap and infinitesimal jackknife variance with Monte Carlo sampling correction, shown in Equation 
\ref{eq:jackknife}, as done in Ling et al. \cite{ling2017high} based on the work of Efron \cite{efron2012model} and Wager 
\cite{wager2014confidence}. 

\begin{equation}
    \sigma_i^2(x)=Cov_j[n_{i,j},t_j(x)]^2+[\bar t_{-_i}(x)-\bar t(x)]^2-ev/T
    \label{eq:jackknife}    
\end{equation}

where $\sigma_i^2(x)$ is the variance at point $x$ due to training point $i$, $n_{i,j}$ is the number of times point $i$ was used to train tree $j$, $\bar t_{-_i}(x)$ is the average of the prediction over the trees that were fit without using point $i$, $\bar t(x)$ average of the prediction over all of the trees,$t_j(x)$ prediction of tree index $j$ for point $x$, $e$ is Euler's number and $v$ is the variance over all trees. Uncertainty estimates are computed as sum of contributions from each estimator for each training point, as seen in Equation \ref{eq:estimator}:

\begin{equation}
    \sigma[M(x)] = \sqrt{ \sum\limits_{i=1}^S max[\sigma^2_i(x), \omega] + \widetilde{\sigma}^{ 2}(x)}
    \label{eq:estimator}
\end{equation}

where $\omega$, the noise threshold, which is equivalent to the $min  \sigma^2(x_i)$, the magnitude of the minimum variance, $[M(x_i)]$ is the model prediction of point $x_i$, $\tilde\sigma^2(x)$ is an explicit bias model, and S is the number of training points. This explicit bias model and uncertainty estimates coincide with those described in Ling et al. \cite{ling2017high}

\subsection{Materials Descriptors and Model Hyperparameters}

For the RF models described in subsection \ref{rf_section} to predict melting temperature, it is vital to represent our alloys with 
descriptors that can serve as inputs. A brute-force approach would use composition as the sole inputs but, given the relative scarcity of 
the data in materials applications, enhancing the inputs with descriptors or features expected to correlate with the outputs is 
highly beneficial.
Previous work by Zhang et al. \cite{zhang2020phase} in which they used a classification algorithm to predict the phases of alloys 
outlined a materials descriptor space and a genetic algorithm strategy to down select a subset of the pool of properties. The complete 
set included 70 properties based on the molar average value and mismatch value for each elemental property of the alloy. Their work 
showed that RFs outperform other models in the prediction task, but their algorithm selects subsets limited to four descriptors. 
Similarly, work by Zhou et al. \cite{zhou2019machine} shows that machine learning models can help the exploration of phase design of 
high entropy alloys. Their model uses 13 descriptors based on the average and standard deviation of properties like atomic radius, 
melting temperature, mixing enthalpy, mixing entropy, electronegativity, bulk modulus and valence electron concentration. 

In this work, atomic properties for elements contained in the alloys are queried from properties available online on 
Pymatgen\cite{pymatgen} and used to generate a unique fingerprint of each alloy using a weight-average rule. The final set of 
descriptors was based on the strongest correlation with melting temperature of our initial set, measured with Pearson correlation 
coefficients. This can be seen in Supplemental Information (SI) as Figure S1. However, it nicely reflects some of the candidates 
proposed by Zhang et al. \cite{zhang2020phase} as it includes properties such as: melting point, electronegativity, boiling 
temperature, atomic mass, atomic radii, and density. Another subset of the highly correlated descriptors relates to mechanical and 
thermal properties and includes Young's modulus, Poisson's ratio, hardness, and coefficient of thermal expansion. Finally, 
electrical resistivity was added to the descriptors as it showed a high correlation to melting temperature.

\begin{figure}[H]
     \centering
     \includegraphics[width=0.6\textwidth]{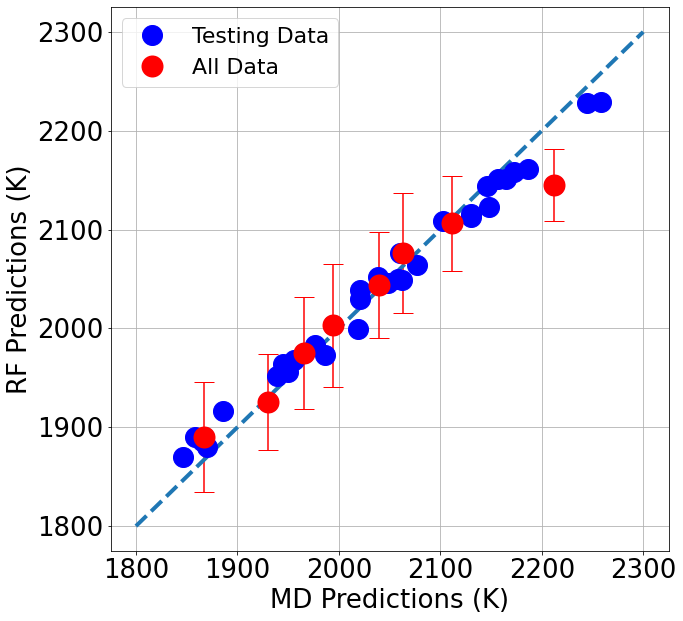}
     \caption{Comparison for MD simulated and RF predicted melting temperatures for the 39 MPCAs compositions in our initial set.} 
     \label{ParityPlot}
\end{figure}

To test our initial model, we split the initial set in an 80/20 train/test split. In Figure \ref{ParityPlot} we present a parity plot 
of the melting temperatures predicted by our RF model and the melting temperatures obtained via MD simulations.

\begin{figure}[H]
     \centering
     \includegraphics[width=0.6\textwidth]{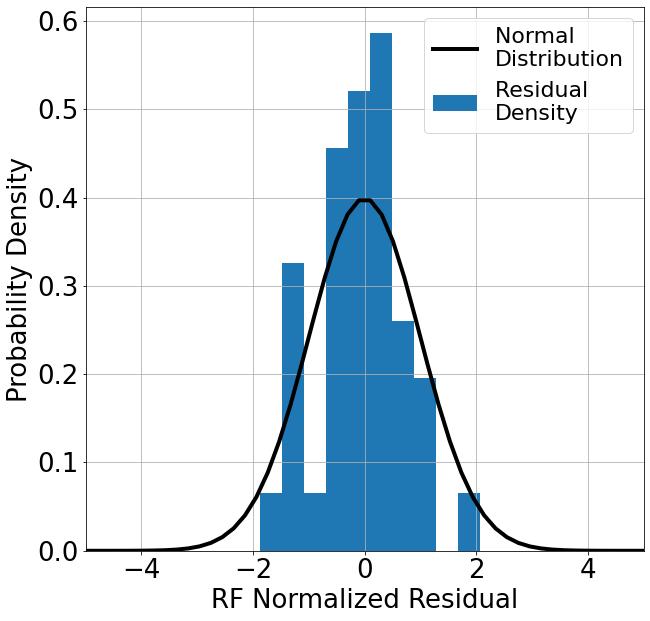}
     \caption{Probability densities of normalized residuals of RF model computed through tenfold cross-validation. Solid line is perfectly calibrated uncertainties.}
     \label{Residual Dist}
\end{figure}

Given the small initial dataset, large deviations in MAE can occur due to the random selection of the testing set. To properly 
determine the model's accuracy, the dataset was shuffled, split, and rerun 30 times through the random forest. The average of the 
MAE for the RF predicted melting temperature with respect to the MD simulated temperature was 36.44 K with an uncertainty estimate of 12.2 K. A 10-fold cross validation analysis was used to create a normalized residual distribution to determine calibration of the uncertainty. Models with well calibrated uncertainties would produce a Gaussian distribution with a mean of zero and unit standard 
deviation as seen in Figure \ref{Residual Dist}.

We optimized the number of estimators (trees) to 350 trees at the maximum depth of each estimator, using the mean absolute error (MAE) as a metric to measure the accuracy of our model predictions. Increasing the number of trees reduces variance, while increasing the maximum depth can help reduce bias. Our optimization stopped based on the idea that the larger quantity of either would improve the RF until we observed diminishing returns. Plots for the optimization of these parameters are included in the supplemental information as Figure S2. 


\subsection{Acquisition functions}
\label{acquisition_section}

Acquisition functions are mathematical expressions designed to select which of the possible tests (MD simulations in our case) to 
carry out next. This is done by considering the model predicted value and uncertainty estimate of the quantity of interest for all 
possible tests. For our case, we use the predicted value and uncertainty estimates provided by the RF model. Various acquisition 
functions have been proposed with different balances between exploitation and exploration. Exploration functions select areas of 
high uncertainty and, on the opposite end, functions based purely on exploitation select candidates with the highest expected values.

A purely exploitative approach would be maximizing the mean predicted value; this maximum mean (MM) function can be written as:

\begin{equation}
    MM:  x^*= argmax \;\; E[M(x_i )]
    \label{eq:MM}
\end{equation}

Such functions can easily get trapped in local maxima, and some degree of exploration is often desirable.
Upper confidence bound (UCB) queries the sample with the maximum predicted value plus its uncertainty estimate. For this study, the adjustable parameter \textit{K} is set to unity (\textit{K} = 1).

\begin{equation}
    UCB:  x^*= argmax \;\; (E[M(x_i )] + K * \sigma[M(x_i )])
    \label{eq:UCB}
\end{equation}

Maximum likelihood of improvement (MLI) chooses the sample with the highest probability of surpassing the current best previously evaluated material. This probability can be computed as the Z-score for predictions that are normally distributed. Therefore, it is represented by the difference between the expected value of the prediction and the value of the current best case, $x_{best}$, over the uncertainty estimate. 

\begin{equation}
    MLI:  x^*= argmax \;\; \frac{E[M(x_i )]-E[M(x_{best} )]}{\sigma[M(x_i )]}
    \label{eq:MLI}
\end{equation}

Maximum expected improvement (MEI) \cite{jones1998efficient} works by modeling our knowledge of the prediction as a normal distribution. It uses the model's mean prediction and uncertainty estimate to draw a normal probability density function at each point. The improvement can then be understood as the probability that the function at this value surpasses our current maximum. Intuitively, this relates to the tail of the distribution crossing our current maximum. In the absence of uncertainties, the function maximum expected improvement (MEI) \cite{jones1998efficient, vazquez2010convergence}, reduces to maximizing the expected values. We note that the MM function was incorrectly labeled as MEI in Refs. \cite{verduzco2021active, ling2017high}.

\begin{equation}
    \begin{split}
        MEI: x^*= argmax \;\; \rho \; ( E[M(x_i )]-E[M(x_{best})], \sigma [M(x_i)] ) \\
        with \;\;
        \rho  \; (z,s) =  \begin{cases} 
          s \phi'(\frac{z}{s}) +
          z \phi(\frac{z}{s}) & s > 0 \\
          max(z,0) & s = 0
        \end{cases}
    \end{split}    
    \label{eq:MEI}
\end{equation}

Finally, maximum uncertainty (MU) is at the extreme end of exploration and only focuses on the candidates with the highest uncertainty in their predictions. MU has little incentive to find the top performer and finds its best use when trying to fine-tune ML models for surrogate-based optimization.

\begin{equation}
    MU:  x^*= argmax \;\; \sigma [M(x_i)]
    \label{eq:MU}
\end{equation}

In these Equations, $x^*$ marks the composition selected by the function to be tested next. $x_i$ are the possible experiments to run (our search space). $x_{best}$ is the current best performer in the training set. $E[M(x_i)]$ is the expectation value of the prediction at point $x$. This expectation value is equal to $\frac{1}{N} \sum\limits_{j}^{n_T} t_j(x)$ where $t_j(x)$ is the prediction of tree $j$ for point $x$ and $n_T$ is the total number of trees. Within MEI, $\phi$ represents the Gaussian cumulative distribution function, and $\phi'$ is the Gaussian probability density function. Finally, ``arg max'' (arguments of the maxima) operation returns the sample material where the function is maximized for the quantity of interest. 

\section{Active learning: exploring high melting temperature alloys}
\label{activelearning_section}

As described above, we started with knowledge of the melting temperature of 39 alloys and use AL to search for the alloy with the highest melting temperature in a 5-dimensional compositional space with 554 possible alloys. The performance of these acquisition functions was assessed from AL workflows with a budget of 40 experiments, mimicking time and resource constraints in real world applications. Here an experiment is equal to the convergence of a composition of an alloy which may require several simulations. As previously described, convergence is defined when the simulation outputs an alloy that reached temperature equilibrium and the fraction of atoms in each of the phases needs to account for 35\% to 65\% of the system.
At each step of the iterative process, we use the RF models to evaluate each acquisition function over all unexplored alloys and select the alloy that maximizes the selected function. A new MD simulation is launched with that alloy and the resulting data is added to the data set and the processes is restarted. 

The MD simulation is run for the desired time and convergence, or lack thereof, is determined with the criteria in Section \ref{MDSims}. Simulation inputs are box length n=18, MD simulation time, and temperatures of the liquid and solid regions. Additional inputs include seeds for pseudo-random number generation for initialization of the initial random structure and atomic velocities. The initial temperatures for the liquid and solid regions from the melting temperature predicted by the RF (T$_{m}$) as follows: T$_{liquid,t=0}$ = 1.25 T$_{m}$ ; T$_{solid, t=0}$ = 0.5 T$_{m}$. If a converged melting temperature calculation is not reached, the initial temperatures are adjusted, and a new simulation launched. The initial temperatures are increased or decreased by 5\% if the resulting structural analysis indicated little liquid or solid left, respectively. This process continues until convergence is achieved. Once convergence is reached the value will be added into the known set, the model will make new predictions, and the process will repeat until our design goal is reached or our experiment budget is exhausted.

\subsection{Active Learning results}

To explore the combination of AL with MD simulations we tested six selection strategies (five information acquisition functions described in Section \ref{acquisition_section} and a random sampling baseline). To assess the effect the noise in the data we considered three simulation times 50, 100 and 200 ps. Each of the 18 runs used the same initial set as described in section \ref{SearchSpace}.

\begin{figure}[H]
     \centering
     \includegraphics[width=.8\textwidth]{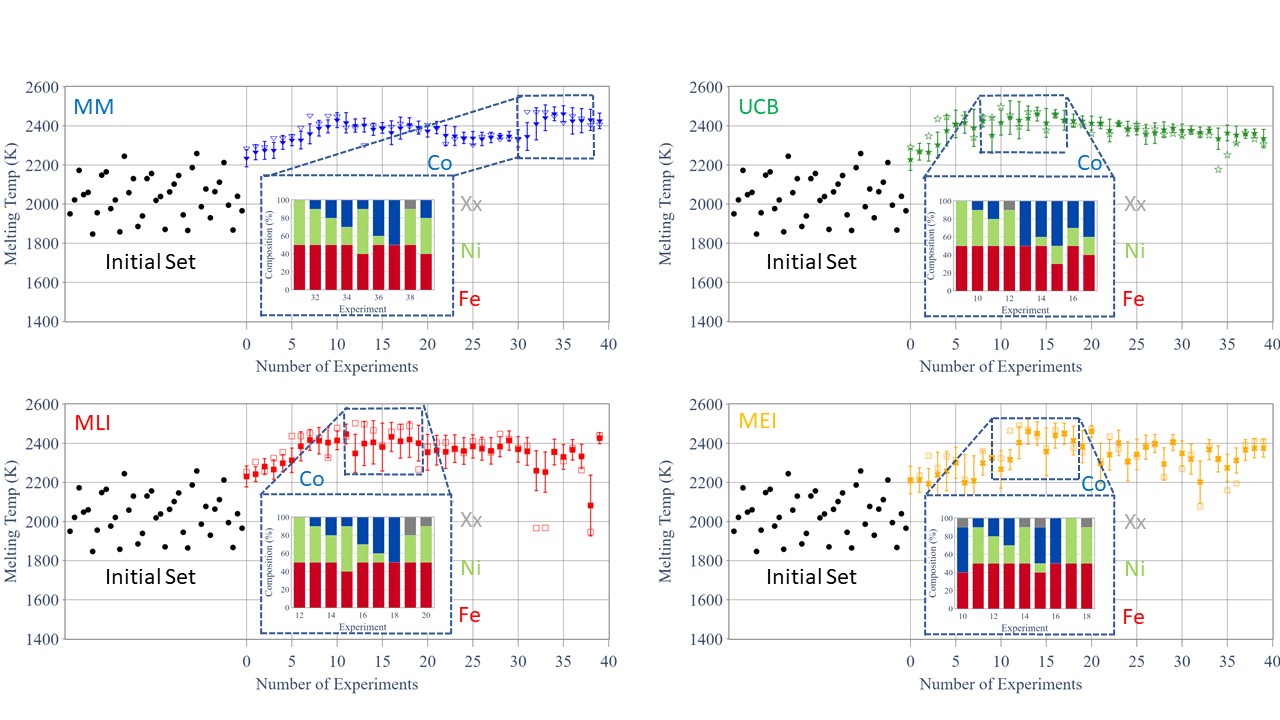}
     \caption{Performance of different acquisition functions in a 40-experiment budget AL run with 50 picoseconds MD simulation time. All functions start from identical initial sets. Open symbols represent MD simulated melting temperatures. The filled symbols with error bars represent RF predicted melting temperatures. Xx represents other elements, Cu and Cr. The insert includes a close-up on the best performing MPCAs compositions. These compositions contain high quantities of Fe.}
     \label{50picoCompActive}
\end{figure}

\begin{figure}[H]
     \centering
     \includegraphics[width=.8\textwidth]{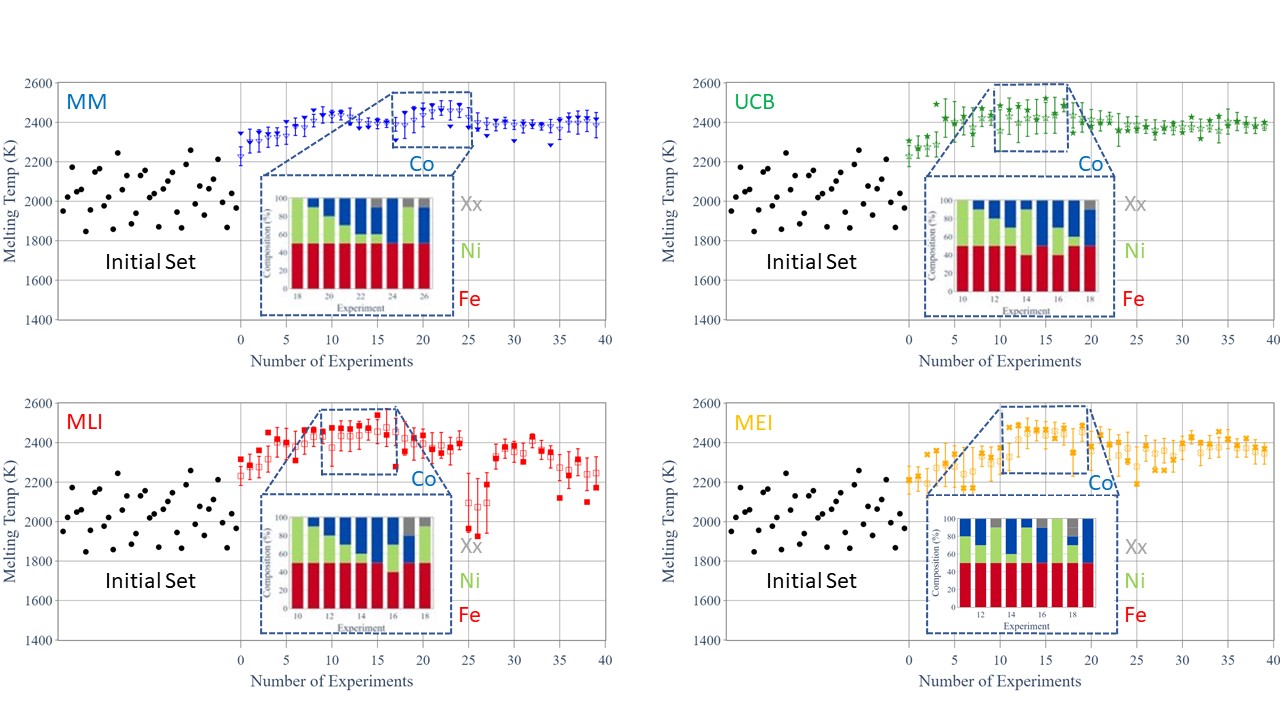}
     \caption{Performance of different acquisition functions in a 40-experiment budget AL run with 100 picoseconds MD simulation time. All functions start from identical initial sets. Open symbols represent MD simulated melting temperatures. The filled symbols with error bars represent RF predicted melting temperatures. Xx represents other elements, Cu and Cr. The insert includes a close-up on the best performing MPCAs compositions. These compositions contain high quantities of Fe.}
     \label{100picoCompActive}
\end{figure}

\begin{figure}[H]
     \centering
     \includegraphics[width=.8\textwidth]{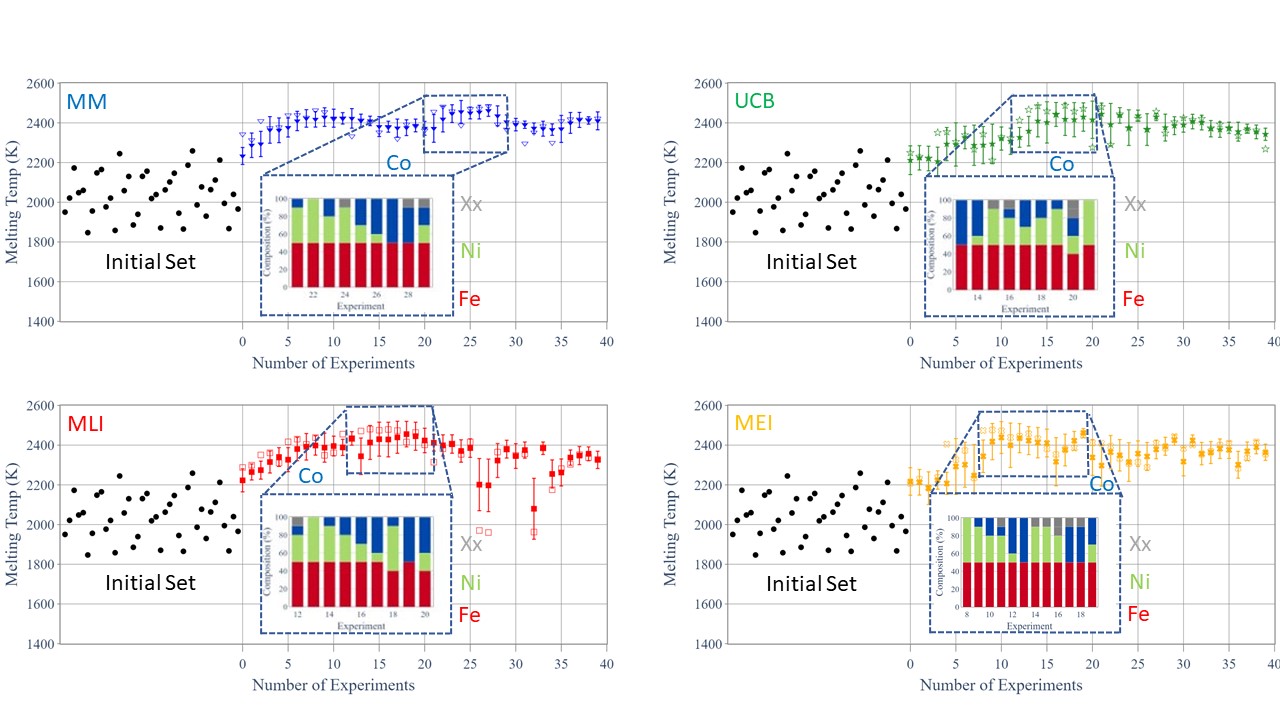}
     \caption{Performance of different acquisition functions in a 40-experiment budget AL run with 200 picoseconds MD simulation time. All functions start from identical initial sets. Open symbols represent MD simulated melting temperatures. The filled symbols with error bars represent RF predicted melting temperatures. Xx represents other elements, Cu and Cr. The insert includes a close-up on the best performing MPCAs compositions. These compositions contain high quantities of Fe.}
     \label{200picoCompActive}
\end{figure}

Figures \ref{50picoCompActive}, \ref{100picoCompActive} and \ref{200picoCompActive} show the melting temperature, both predicted by the RF (filled symbols) and the one obtained from the subsequent MD simulation (open symbols), as a function of experiment number for the various acquisition functions and simulation times. The insets in Figures \ref{50picoCompActive}, \ref{100picoCompActive} and \ref{200picoCompActive}, show the composition of the alloys with the highest predicted melting temperatures. Tables \ref{50_picosec_table}, \ref{100_picosec_table} and \ref{200_picosec_table} summarize the various AL outcomes: for the alloy with the highest melting temperature within the budget of 40 possible compositions for each case, we list the temperature predicted by the RF and the actual value obtained from the MD simulations, the uncertainty in the RF model, number of experiments required to locate the alloy with the highest melting temperature and its composition. 


\begin{table}[H]
\caption{Best results from information acquisition functions running simulations for 100 picoseconds. RF prediction melting temperature and uncertainty taken when best composition selected for each acquisition function. }
\label{50_picosec_table}
\vskip 0.15in
\begin{center}
\begin{small}
\begin{sc}
\resizebox{\textwidth}{!}{%
\begin{tabular}{lcccccr}
\toprule
\thead{Acquisition \\ function} & \thead{MD simulated \\ Tm (K)} &  \thead{RF predicted \\ Tm (K)} & \thead{RF \\ uncertainty (K)} & Experiment & Composition  \\
\midrule
MM & 2472  & 2345 & 71 & 31 & Fe 50\%, Ni 50\%   \\
UCB & 2498 & 2416 & 63 & 10 & Co 10\%, Fe 50\%, Ni 40\% \\
MLI & 2502  & 2348 & 98 & 12 & Fe 50\%, Ni 50\% \\
MEI & 2492 & 2403 & 82 & 12 & Co 20\%, Fe 50\%, Ni 30\% \\
MU  & 2282 & 2137 & 118 & - & Co 40\%, Fe 20\%, Ni 40\% \\
RAND  & 2382 & 2202 &  52 & - & Cu 10\%, Fe 50\%, Ni 40\%    \\

\bottomrule
\end{tabular}}
\end{sc}
\end{small}
\end{center}
\vskip -0.1in
\end{table}

\begin{table}[H]
\caption{Best results from information acquisition functions running simulations for 100 picoseconds. RF prediction melting temperature and uncertainty taken when composition selected selected for each acquisition function.}
\label{100_picosec_table}
\vskip 0.15 in
\begin{center}
\begin{small}
\begin{sc}
\resizebox{\textwidth}{!}{%
\begin{tabular}{lcccccr}
\toprule
\midrule
\thead{Acquisition \\ function} & \thead{MD simulated \\ Tm (K)} &  \thead{RF predicted \\ Tm (K)} & \thead{RF \\ uncertainty (K)} & Experiment & Composition  \\
\midrule
MM & 2489  & 2460 & 36 & 24 & Fe 50\%, Ni 50\%   \\
UCB & 2521 & 2422 & 78 & 15 & Fe 50\%, Ni 50\% \\
MLI & 2539  & 2455 & 60 & 15 & Fe 50\%, Ni 50\% \\
MEI & 2486 & 2415 & 87 & 12 & Co 30\%, Fe 50\%, Ni 20\% \\
MU  & 2299 & 2160 & 117 & - & Co 20\%, Cu 20\%, Fe 50\%, Ni 10\% \\
RAND  & 2382 & 2202 & 52 & - & Cu 10\%, Fe 50\%, Ni 40\% \\
\bottomrule
\end{tabular}}
\end{sc}
\end{small}
\end{center}
\vskip -0.1in
\end{table}

\begin{table}[H]
\caption{Best results from information acquisition functions running simulations for 200 picoseconds. RF prediction melting temperature and uncertainty taken when composition selected selected for each acquisition function. }
\label{200_picosec_table}
\vskip 0.15in
\begin{center}
\begin{small}
\begin{sc}
\resizebox{\textwidth}{!}{%
\begin{tabular}{lcccccr}
\toprule
\thead{Acquisition \\ function} & \thead{MD simulated \\ Tm (K)} &  \thead{RF predicted \\ Tm (K)} & \thead{RF \\ uncertainty (K)} & Experiment & Composition  \\
\midrule
MM & 2485  & 2461 & 29 & 27 & Fe 50\%, Ni 50\%   \\
UCB & 2487 & 2417 & 81 & 17 & Co 30\%, Fe 50\%, Ni 20\% \\
MLI & 2480  & 2413 & 85 & 14 & Co 10\%, Fe 50\%, Ni 40\% \\
MEI & 2476 & 2418 & 91 & 9 & Co 10\%, Fe 50\%, Ni 50\% \\
MU  & 2285 & 2160 & 118 & - & Co 40\%, Cu 10\%, Fe 30\%, Ni 20\% \\
RAND  & 2432 & 2360 & 80 & - & Co 40\%, Fe 40\%, Ni 20\% \\
\bottomrule
\end{tabular}}
\end{sc}
\end{small}
\end{center}
\vskip -0.1in
\end{table}

\begin{figure}[H]
     \centering
     \includegraphics[width=.5\textwidth]{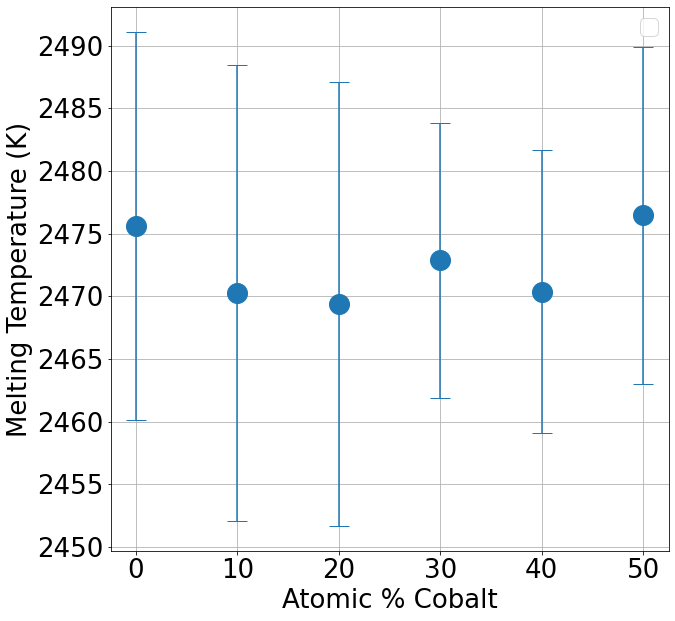}
     \caption{Predicted melting temperature best performing alloys with composition Fe$_{50}$Co$_{x}$Ni$_{50-x}$. Mean values and uncertainty estimates over 36 independent runs are shown.}
     \label{BestAlloys}
\end{figure}

Quite interestingly, all acquisition functions that take exploitation into consideration (MLI, UCB, MEI, and MM) find the alloys with the highest melting temperatures in few experiments, regardless of the simulation time used. This is quite remarkable given the significant variability in output for the 50 ps long simulations. We find that highest melting temperature alloys consist of 50 at.\% Fe with the remaining 50 at.\% distributed between Ni and Co; this set of alloys are predicted to have nearly identical melting temperatures, see Figure \ref{BestAlloys}. As expected, the MU function and random exploration do not explore high-melting temperature
compositions within the 40 experiment budget. The graph seen for MM, UCB, MLI, and MEI has been made for MU and random and are in the SI as S3, S4. 

In all our tests, we find that the acquisition functions that combine both exploration and exploitation (UCB, MLI, MEI) consistently outperform the purely exploitative function (MM), which can spend resources in local maxima. As seen in Tables \ref{50_picosec_table}-\ref{200_picosec_table} UCB, MLI, and MEI find the optimal composition within less than 17 experiments while MM found it past 27. While in our example MM finds the family of best performer candidates, it takes approximately twice as many experiments as the other functions.

\subsection{Noisy data and RF uncertainty}
\label{sect:NoisyDataRF}

We find that the AL workflow can find best performer alloys in approximately the same number of iterations regardless of 
the simulation time use. Between 9 and 17 simulations were required for the UCB, MEI, and MLI strategies. As mentioned above, this 
is despite the significant level of noise, especially in the 50 ps runs.  We also find that while longer MD simulations do not result 
in a reduction in the number of experiments required to achieve the design goal, it results in improved accuracy of the RF prediction. 
Tables \ref{50_picosec_table}, \ref{100_picosec_table} and \ref{200_picosec_table} show that the workflow with 50 ps runs significantly 
underestimates the temperature of the optimal alloy and the uncertainty estimates are overly optimistic.  

\begin{figure}[H]
     \centering
     \includegraphics[width=\textwidth]{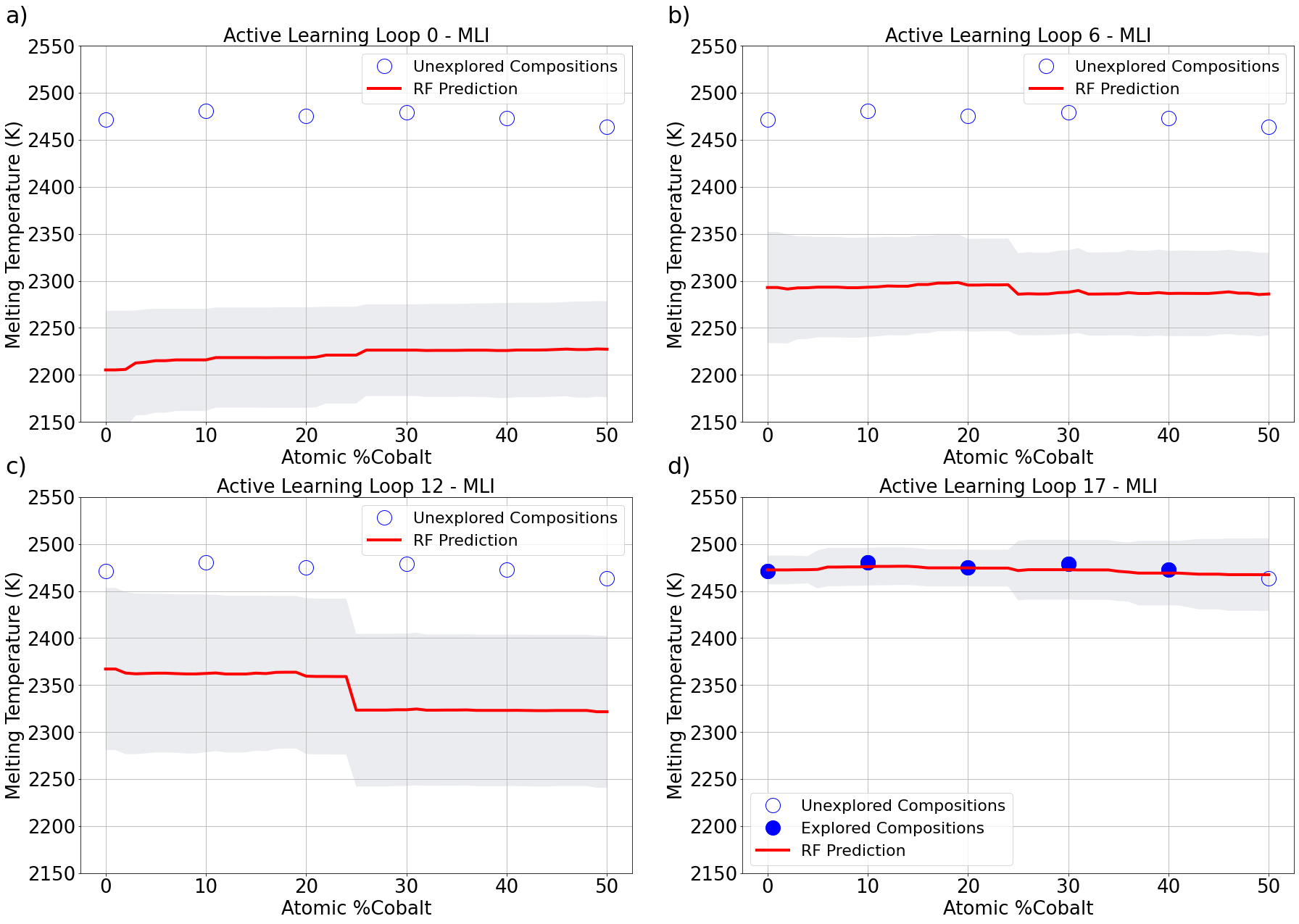}
     \caption{Red lines represent the predicted mean melting temperature and shaded region represents the uncertainty estimates for predictions on compositions as a function of Co content in Fe$_{50}$Co$_{x}$Ni$_{50-x}$ at various stages of the AL workflow. Whereby x starts at 0\% a goes to 50\% with 1\% step size. Results correspond to the MLI function with a 50 ps MD simulation time. Open circles represent MD-simulated values unknown to the model at the time and filled symbols represent the values included in the model.}
     \label{RandomForestTreeError}
\end{figure}

To better understand how the models gain knowledge in the presence of noise we explore the evolution of the model predictions for 
Fe$_{50}$Co$_{x}$Ni$_{50-x}$ alloys, the top performers, at various stages during the AL process. Figure \ref{RandomForestTreeError} 
shows mean and uncertainty estimate predictions at four stages of an MLI workflow for the 50 ps case; results for 100 and 200 ps 
simulations show similar trends and are included in the SI as Figures S5, S6.
Open circles denote MD simulations not already explored at the corresponding cycle and filled circles represent the same values 
for compositions that have been explored.
We note that noise in the data affects both mean predictions and uncertainty estimates and can thus have non-trivial effects on 
AL decisions. Figure \ref{RandomForestTreeError} shows that even with only the initial 39 datapoints (Iteration 0), the model predicts 
relatively high melting temperature for these alloys, comparable to the highest melting temperature materials in the initial set, see 
Figure \ref{50picoCompActive}. As the AL workflow with MLI explore compositions, the mean prediction for the selected family improves from \ref{RandomForestTreeError} (a) to (b) as the mean moves towards the true MD temperatures of the alloys. 
At step 12, Fig. \ref{RandomForestTreeError} (d), the model predicts high melting temperatures for the alloys with x=0, 10, 20 \% and
x=0 is selected for simulation. In subsequent steps, the MLI algorithm hones in the rest of the family and improves the model accuracy. 
When comparing the results for 50 ps runs with those of 100 and 200 ps (SI Figures S5 and S6) we observe similar paths to the 
optimal alloys with a decrease in uncertainties for longer simulation times. This is because the variability in the MD simulations
is relatively small compared to the differences in melting temperatures across alloys.

\section{Conclusions}
\label{conclusion_section}

The combination of machine learning and physics-based simulations holds great promise to guide the experimental search and accelerate the development of novel materials. The use of surrogate-based models that can make use of computational simulations instead of experiments can save resources and time in design cycles. We find the molecular dynamics simulations, even under significant stochastic noise, can be a powerful tool to be used in active learning efforts.

We developed an AL workflow with MD simulations for the discovery of MPCAs materials for high-temperature applications. Starting from a limited initial set, we created a set of descriptors and a RF model to use as a surrogate for the optimization of our AL model. We determined that RF models are capable of handling multi-dimensional spaces with a limited initial dataset. We paired our RF model with a cloud-based simulation environment for automatic querying of melting temperatures from MD simulations with an interatomic potential for FCC alloys. We analyzed the effect of MD simulation time in the overall uncertainty quantification when running such calculations and showed that it has an important effect in capturing the sample-to-sample variability caused by the stochastic nature of MD. We compared the performance of four different acquisition functions in a closed, but high-dimensional, search space. We found that the best performing acquisition functions, MLI, MEI, and UCB, are combinations of exploitation and exploration.

The workflow we developed can be adapted toward the optimization of other material properties and to the use of other sources of information other than computational simulations. We provide the analysis and all other relevant information in a public SimTool on nanoHUB.org.

\section{Models and Data Availability}
\label{availability_section}

The entire simulation workflow to calculate melting temperature, including pre- and post-analysis was implemented as a SimTool 
\cite{desai2020introduction} in nanoHUB \cite{strachan2010cyber} and is available for online simulation \cite{nh_meltheas}. 
SimTools \cite{desai2020introduction} are complete simulation workflows with verified inputs and outputs that enable workflows 
and the data they generate to be findable, accessible, and reproducible. An additional advantage of using SimTools is that every simulation performed in nanoHUB is cached, so that given identical input parameters, cached results are retrieved rather than re-running a simulation. The stand-alone SimTool \cite{nh_meltheas} and the AL workflow are available for online simulations in nanoHUB \cite{nh_workflow_activemeltheas}.

\section*{Acknowledgements}

This effort was supported by the US National Science Foundation (DMREF-1922316). We acknowledge computational resources from nanoHUB and Purdue University through the Network for Computational Nanotechnology. J. C. V. thanks the Science and Technology Council of M{\'e}xico (Consejo Nacional de Ciencia y Tecnología, CONACYT) for partial financial support of this research. 

\section*{Ethics declarations}

\subsection*{Conflict of interest}

On behalf of all authors, the corresponding author states that there is no conflict of interest.

\subsection*{Author Information}

\textbf{Corresponding Author}

\textbf{Alejandro Strachan} - School of Materials Engineering and Birck Nanotechnology Center, Purdue University, West Lafayette, Indiana 47907; Email: strachan@purdue.edu \\

\bibliographystyle{unsrt}
\bibliography{references.bib}

\end{document}